\documentclass[preprint,tightenlines,aps]{revtex4}
\begin{document}
\title{Open-charm meson spectroscopy}
\author{J. Vijande, F. Fern\'{a}ndez, A. Valcarce}
\affiliation{Grupo de F\'{\i}sica Nuclear and IUFFyM,
Universidad de Salamanca, E-37008 Salamanca,
Spain}
\vspace*{1cm} 
\begin{abstract}
We present a theoretical framework that accounts for 
the new $D_J$ and $D_{sJ}$ mesons measured in 
the open-charm sector. These resonances are 
properly described if considered as a mixture
of conventional $P-$wave quark-antiquark 
states and four-quark components. 
The narrowest states are basically $P-$wave quark-antiquark
mesons, while the dominantly four-quark states are shifted above 
the corresponding two-meson threshold, being broad 
resonances. We study the electromagnetic decay widths
as basic tools to scrutiny their nature. 
The proposed explanation incorporates in a 
natural way the most recently discovered mesons 
in charmonium spectroscopy.

\vspace*{2cm} 
\noindent Keywords: Charm mesons, Charm-strange mesons,  quark models.
\newline
\noindent Pacs: 14.40.Lb, 14.40.Ev, 12.39.Pn. 
\end{abstract}
\maketitle
\newpage

\section{INTRODUCTION}

During the last few years, heavy meson spectroscopy is living a 
continuous excitation due to the discovery of several new charmed 
mesons. Two years ago BABAR Collaboration at the 
Standford Linear Accelerator Center (SLAC) reported the observation 
of a charm-strange state, the $D_{sJ}^*(2317)$ \cite{Bab03}. It was confirmed by
CLEO Collaboration at the Cornell Electron Storage Ring \cite{Cle03} and 
also by Belle Collaboration at KEK \cite{Bel04}.
Besides, BABAR had also pointed out to the existence 
of another charm-strange meson, the $D_{sJ}(2460)$ \cite{Bab03}. This
resonance was measured by CLEO \cite{Cle03} and 
confirmed by Belle \cite{Bel04}. 
Belle results \cite{Bel04} are consistent with the 
spin-parity assignments of $J^P=0^+$ for the 
$D^*_{sJ}(2317)$ and $J^P=1^+$ for the $D_{sJ}(2460)$.
Thus, these two states are definitively well established, 
confirmed independently by different experiments. 
They present unexpected properties, quite different from
those predicted by quark potential models. If they would correspond 
to standard $P-$wave mesons made of a charm quark, $c$, and a
strange antiquark, $\overline{s}$, 
their masses would be larger \cite{God91},
around 2.48 GeV for the $D_{sJ}^*(2317)$ and 2.55 GeV 
for the $D_{sJ}(2460)$. They would be therefore above the
$DK$ and $D^*K$ thresholds, respectively, being broad resonances.
However the states observed by BABAR and CLEO are very
narrow, $\Gamma < 4.6$ MeV for the $D_{sJ}^*(2317)$ 
and $\Gamma < 5.5$ MeV for the $D_{sJ}(2460)$. 

The intriguing situation of the charm-strange mesons 
has been translated to the nonstrange sector with the Belle 
observation \cite{Belb4} of a nonstrange broad scalar resonance, $D^*_0$,
with a mass of $2308\pm 17\pm 15\pm 28$ MeV/c$^2$
and a width $\Gamma=276 \pm 21 \pm 18 \pm 60$ MeV.
A state with similar properties has been suggested by 
FOCUS Collaboration at Fermilab \cite{Foc04} during the 
measurement of masses and widths of excited charm mesons
$D^*_2$. They found significant evidence for a broad excess
parametrized as an $S-$wave resonance. Without being able
to clearly distinguish the origin of this broad excess,
they conclude that their results are in agreement with
the Belle results \cite{Belb4}. This state generates 
for the open-charm nonstrange mesons a very similar
problem to the one arising in the strange sector with the
$D_{sJ}^*(2317)$. If the $D^*_0(2308)$  would correspond 
to a standard $P-$wave meson made of a charm quark, $c$, and a
light antiquark, $\overline{n}$, its mass would have to be larger,
around 2.46 GeV. In this case, the quark potential models prediction and
the measured resonance are both above the $D\pi$ 
threshold, the large width observed being expected although not
its low mass.

The last step in this series of discoveries has been the
observation of a new charm-strange meson, $D_{sJ}$, announced by 
SELEX Collaboration at Fermilab \cite{Sel04} with a mass of 
$2632.5\pm 1.7$ MeV/c$^2$ and a small width, $\Gamma < 17$ MeV. 
However, up to now no other experiment has been able to confirm
the existence of this resonance \cite{Bab04}. 

There have been many theoretical interpretations for the 
masses and widths of the new resonances, but 
most part of them have been devoted to explain the strange 
states. Ref. \cite{Bev03} made use of a 
unitarized meson model, the existence of a quasi-bound 
$c\overline{s}$ state due to the coupling with the
nearby $S-$wave $DK$ threshold. 
The coupling to the $DK$ channel within a QCD string model an
a chiral Lagrangian was used in Ref. \cite{Sim04}.
Ref. \cite{Cah03} proposed a $c\overline{s}$ 
structure with modified non-central forces. Ref. \cite{Bar03} 
combined HQET with a chiral effective Lagrangian to interpret these
states as the missing $j=1/2$ member of the $c\overline{s}$ $L=1$ ground
state multiplet, being $j$ the angular momentum of the strange quark. The
smaller mass of the $D_{sJ}^*(2317)$ is attributed in Ref. \cite{Hwa04} 
to a coupled channel effect. 
Finally more involved qualitative solutions like 
$DK-$molecules \cite{Barb3}, $D_s\pi-$atoms \cite{Szc03}, four-quark states 
\cite{Che03,Cheb3} and their combination 
with $q\overline{q}$ states \cite{Bro04} 
have been invoked to explain why the new states have a nature so different
from canonical $q\overline{q}$ mesons. 
Although the nonstrange partners of the $D_{sJ}$'s
have received much less attention, as discussed above,
they present similar spectroscopic properties that should 
be acknowledge altogether
with those of the strange states in any reliable model.

The difficulties to identify the $D_J$ and $D_{sJ}$ states with 
conventional $c\overline{q}$ mesons are rather similar to those 
appearing in the light-scalar meson sector \cite{Bal03} and may be indicating
that other configurations are playing a role, as could be 
for example four-quark contributions.
$q\overline{q}$ states are more easily identified with physical hadrons 
when virtual quark loops are not important. 
This is the case of the pseudoscalar 
and vector mesons, mainly due to the $P-$wave nature of this hadronic dressing. 
On the contrary, in the scalar sector 
it is the $q\overline{q}$ pair the one in
a $P-$wave state, whereas quark loops may be in a $S-$wave. 
In this case the intermediate hadronic states that are created 
may play a crucial role in the composition of the resonance,
in other words unquenching may be important. The vicinity of these components
to the lightest $q\bar q$
state implies that they have to be considered. This has been shown as a
possible interpretation of the low-lying light-scalar mesons, where the
coupling of the scalar $q\overline{q}$ nonet 
to the lightest $qq\bar{q} \bar{q}$ 
configurations allows for an almost one-to-one correspondence
between theoretical states and experiment \cite{Vij05}. 

In this work we pretend to explore the same ideas for the
understanding of the properties of the 
$D_J$ and $D_{sJ}$ meson states. For this purpose, in the next section we will
present our calculating scheme and the basic ingredients of the
constituent quark model used. Section III will be devoted to present and discuss
our results in connection with those obtained for the light-scalar mesons.
Finally, in Sec. IV we will resume the most important conclusions of our work.

\section{CALCULATING FRAMEWORK}

In non-relativistic quark models gluon degrees 
of freedom are frozen and therefore the wave 
function of a zero baryon number (B=0) hadron may be written as
\begin{equation}
\label{mes-w}
\left|\rm{B=0}\right>=\Omega_1\left|q\bar q\right>+\Omega_2\left|qq\bar q \bar q\right>+....
\end{equation}
where $q$ stands for quark degrees of 
freedom and the coefficients $\Omega_i$ take into 
account the mixing of four-quark and $q \bar q$ states.
$\left|\rm{B=0}\right>$ systems could then be 
described in terms of a hamiltonian 
\begin{equation}
H=H_0 + H_1 \,\,\,\,\, {\rm being} \,\,\,\,\,  
H_0 = \left( \matrix{H_{q\bar q} & 0 \cr
0 & H_{qq\bar q\bar q} \cr } \right) \,\,\,\,
H_1 = \left( \matrix{0 & V_{q\bar q \leftrightarrow qq\bar q\bar q} \cr 
V_{q\bar q \leftrightarrow qq\bar q\bar q} & 0 \cr } \right)\,,
\label{eq1}
\end{equation}
where $H_0$ is a constituent quark model hamiltonian described below and
$H_{1}$, that takes into account the mixing between
$q\overline{q}$ and $qq\bar{q}\bar{q}$ configurations, includes the
annihilation operator of a quark-antiquark pair into the vacuum. This
operator could be described using the $^{3}P_{0}$ model, however, since this 
model depends on the vertex parameter, we prefer in a
first approximation to parametrize this coefficient 
by looking to the quark pair that is annihilated and 
not to the spectator quarks that will form the final 
$q\overline{q}$ state. Therefore we have taken 
$V_{q\overline{q} \leftrightarrow qq\bar{q}\bar{q}}=\gamma $.
If this coupling is weak enough one can solve independently 
the eigenproblem for the hamiltonians $H_{q\overline{q}}$ 
and $H_{qq\bar{q}\bar{q}}$, treating $H_{1}$ perturbatively. 
To ensure that the perturbative treatment is justified, 
$\gamma$ cannot take all possible values, being restricted to
$|\gamma/(E^{\,n}_{J^{PC}}-E^{\,n+1}_{J^{PC}})|^2\leq1$. This restriction
will limit the energy range of the mixed states once
the unmixed energies are calculated.
The two-body problem has been solved exactly by means of the Numerov
algorithm \cite{Vijb5}. The four-body problem has been solved
by means of a variational method using the most general 
combination of gaussians as trial
wave functions \cite{Suz98,Veu04}. 
In particular, the so-called {\it mixed terms} (mixing the various
Jacobi coordinates) that are known to have a great influence in
the light quark case have been considered. 

Although the constituent quark model used is described in 
Ref. \cite{Vijb5}, let us outline here its basic ingredients.
Since the origin of the quark model hadrons have been considered to be
built by constituent (massive) quarks. Nowadays it is widely
recognized that the constituent quark mass of light quarks appears because 
of the spontaneous
breaking of the original chiral symmetry of the QCD Lagrangian, what 
gives rise to boson-exchange interactions between quarks.
The different terms of the potential
contain central and tensor or central and spin-orbit contributions
that will be grouped for consistency. Therefore, the chiral
part of the quark-quark interaction can be resumed as follows,
\begin{equation}
V_{\chi}(\vec r_{ij}) \, = \,
V_{\chi}^C(\vec r_{ij}) \, + \,
V_{\chi}^T(\vec r_{ij}) \, + \,
V_{\chi}^{SO}(\vec r_{ij}) \, ,
\label{teq}
\end{equation}
where $C$ stands for central, $T$ for tensor, and $SO$ for spin-orbit
potentials. The central part of the quark-quark meson-exchange potentials are given by:
\begin{equation}
V^C_{\chi}(\vec r_{ij}) \, = \,
V^C_{\pi}(\vec r_{ij}) \, + \,
V^C_{\sigma}(\vec r_{ij}) \, + \,
V^C_{K}(\vec r_{ij}) \, + \,
V^C_{\eta}(\vec r_{ij}) \, ,
\end{equation}
each contribution given by,
\begin{eqnarray}
V^C_{\pi}(\vec{r}_{ij})&=&{\frac{g_{ch}^{2}}{{4\pi }}}{\frac{m_{\pi}^{2}}{{\
12m_{i}m_{j}}}}{\frac{\Lambda _{\pi}^{2}}{{\Lambda _{\pi}^{2}-m_{\pi}^{2}}}}%
m_{\pi}\left[ Y(m_{\pi}\,r_{ij})-{\frac{\Lambda _{\pi}^{3}}{m_{\pi}^{3}}}%
Y(\Lambda _{\pi}\,r_{ij})\right] (\vec{\sigma}_{i}\cdot \vec{\sigma}%
_{j})\sum_{a=1}^{3}{(\lambda _{i}^{a}\cdot \lambda _{j}^{a})}\,, \nonumber \\
V^C_{\sigma}(\vec{r}_{ij})&=&-{\frac{g_{ch}^{2}}{{4\pi }}}
{\frac{\Lambda_{\sigma}^{2}%
}{{\ \Lambda _{\sigma}^{2}-m_{\sigma}^{2}}}}m_{\sigma}
\left[ Y(m_{\sigma}\,r_{ij})-{\frac{%
\Lambda _{\sigma}}{m_{\sigma}}} 
Y(\Lambda _{\sigma}\,r_{ij})\right]\, , \label{zp}\\
V^C_{K}(\vec{r}_{ij})&=&{\frac{g_{ch}^{2}}{{4\pi }}}{\frac{m_{K}^{2}}{{\
12m_{i}m_{j}}}}{\frac{\Lambda _{K}^{2}}{{\Lambda _{K}^{2}-m_{K}^{2}}}}m_{K}%
\left[ Y(m_{K}\,r_{ij})-{\frac{\Lambda _{K}^{3}}{m_{K}^{3}}}Y(\Lambda
_{K}\,r_{ij})\right] (\vec{\sigma}_{i}\cdot \vec{\sigma}_{j})\sum_{a=4}^{7}{%
(\lambda _{i}^{a}\cdot \lambda _{j}^{a})}\, , \nonumber \\
V^C_{\eta }(\vec{r}_{ij})&=&{\frac{g_{ch}^{2}}{{4\pi }}}{\frac{m_{\eta }^{2}%
}{{\ 12m_{i}m_{j}}}}{\frac{\Lambda _{\eta }^{2}}{{\Lambda _{\eta
}^{2}-m_{\eta }^{2}}}}m_{\eta }\left[ Y(m_{\eta }\,r_{ij})-{\frac{\Lambda
_{\eta }^{3}}{m_{\eta }^{3}}}Y(\Lambda _{\eta }\,r_{ij})\right] (\vec{\sigma}%
_{i}\cdot \vec{\sigma}_{j})\left[ cos\theta_P(\lambda _{i}^{8}\cdot
\lambda _{j}^{8})-sin\theta _P\right] \, , \nonumber
\end{eqnarray}
the angle $\theta_P$ appears as a consequence of considering
the physical $\eta$ instead the octet one. 
$g_{ch}=m_{q}/f_{\pi }$, the $\lambda ^{\prime }s$ are the $SU(3)$
flavor Gell-Mann matrices and the $\sigma$'s are the spin quark Pauli matrices.
$m_{i}$ is the quark
mass and $m_{\pi}$, $m_{K}$ and $m_{\eta }$ are the masses of the $SU(3)$
Goldstone bosons, taken to be their experimental values. 
The $\Lambda_i$'s are cutoff parameters. $m_{\sigma}$ is
determined through the PCAC relation 
$m_{\sigma}^{2}\sim m_{\pi}^{2}+4\,m_{u,d}^{2}$.
Finally, $Y(x)$ is the standard Yukawa function defined by $Y(x)=e^{-x}/x$.

There are three different contributions to the tensor potential,
\begin{equation}
V_{qq}^T(\vec r_{ij}) \, = \,
V_{\pi}^T(\vec r_{ij}) \, + \,
V_{K}^{T}(\vec r_{ij}) \, + \,
V_{\eta}^T(\vec r_{ij}) \, ,
\end{equation}
each term given by,
\begin{eqnarray}
V_{\pi}^{T}(\vec{r}_{ij})&=&{\frac{g_{ch}^{2}}{{4\pi }}}{\frac{m_{\pi}^{2}}{{%
12m_{i}m_{j}}}}{\frac{\Lambda _{\pi}^{2}}{{\Lambda _{\pi}^{2}-m_{\pi}^{2}}}}%
m_{\pi}\left[ H(m_{\pi}\,r_{ij})-{\frac{\Lambda _{\pi}^{3}}{m_{\pi}^{3}}}%
H(\Lambda _{\pi}\,r_{ij})\right] 
S_{ij}\sum_{a=1}^{3}{(\lambda _{i}^{a}\cdot \lambda _{j}^{a})}\, , \nonumber \\
V_{K}^{T}(\vec{r}_{ij})& =&{\frac{g_{ch}^{2}}{{4\pi }}}{\frac{m_{K}^{2}}{{%
12m_{i}m_{j}}}}{\frac{\Lambda _{K}^{2}}{{\Lambda _{K}^{2}-m_{K}^{2}}}}m_{K}%
\left[ H(m_{K}\,r_{ij})-{\frac{\Lambda _{K}^{3}}{m_{K}^{3}}}H(\Lambda
_{K}\,r_{ij})\right] 
S_{ij}\sum_{a=4}^{7}{(\lambda _{i}^{a}\cdot \lambda _{j}^{a})}\, , \\
V_{\eta}^{T}(\vec{r}_{ij})& =&{\frac{g_{ch}^{2}}{{4\pi }}}{\frac{m_{\eta
}^{2}}{{12m_{i}m_{j}}}}{\frac{\Lambda _{\eta }^{2}}{{\Lambda _{\eta
}^{2}-m_{\eta }^{2}}}}m_{\eta }\left[ H(m_{\eta }\,r_{ij})-{\frac{\Lambda
_{\eta }^{3}}{m_{\eta }^{3}}}H(\Lambda _{\eta }\,r_{ij})\right] 
S_{ij}\left[ cos\theta_P(\lambda _{i}^{8}\cdot \lambda
_{j}^{8})-sin\theta_P\right] \, , \nonumber
\end{eqnarray}
being $S_{ij} \, = \, 3 \, ({\vec \sigma}_i \cdot
{\hat r}_{ij}) ({\vec \sigma}_j \cdot  {\hat r}_{ij})
\, - \, {\vec \sigma}_i \cdot {\vec \sigma}_j$
the quark tensor operator and $H(x)=(1+3/x+3/x^2)\,Y(x)$.

Finally, the spin-orbit potential only presents a contribution
coming form the scalar part of the interaction, 
\begin{equation}
V_{qq}^{SO}(\vec{r}_{ij}) = 
V_{\sigma}^{SO}(\vec{r}_{ij}) =-{\frac{g_{ch}^{2}}{{4\pi }}}{\frac{\Lambda
_{\sigma}^{2}}{{\Lambda _{\sigma}^{2}-m_{\sigma}^{2}}}}{\frac{m_{\sigma}^{3}}
{{2m_{i}m_{j}}}}%
\left[ G(m_{\sigma}\,r_{ij})-{\frac{\Lambda_{\sigma}^{3}}{m_{\sigma}^{3}}}
G(\Lambda_{\sigma}\,r_{ij})\right] \vec{L}\cdot \vec{S} 
\end{equation}
where $G(x)=(1+1/x)\,Y(x)/x$. 

QCD perturbative effects are taken into account
through the one-gluon-exchange (OGE) potential \cite{Ruj75}.
The nonrelativistic reduction of the one-gluon-exchange diagram in QCD for point-like
quarks presents a contact term that, when not treated perturbatively, 
leads to collapse \cite{Bha80}. This is why one
maintains the structure of the OGE, but the $\delta$ function is
regularized in a suitable way. This regularization, justified by 
the finite size of the systems studied, has to be
flavor dependent \cite{Wei83}. As a consequence, the central part of the
OGE reads,
\begin{equation}
V^C_{OGE}(\vec{r}_{ij}) ={\frac{1}{4}}\alpha _{s}\,\vec{\lambda ^{c}}%
_{i}\cdot \vec{\lambda^{c}}_{j}\,\left\{ {\frac{1}{r_{ij}}}-{\frac{1}{%
6m_{i}m_{j}}}\vec{\sigma}_{i}\cdot \vec{\sigma}_{j}
\,{\frac{{e^{-r_{ij}/r_{0}(\mu )}}}{r_{ij}\,
r_0^2(\mu)}}\right\} \, ,
\end{equation}
where $\lambda^{c}$ are the $SU(3)$ color matrices, 
$\alpha_s$ is the quark-gluon coupling constant, and
$r_0(\mu)=\hat r_0 \mu_{nn}/\mu_{ij}$, where $\mu_{ij}$ is the 
reduced mass of quarks $ij$ ($n$ stands for the light $u$ and $d$
quarks) and $\hat r_0$ is a parameter to be determined from the data.

The noncentral terms of the OGE present a similar problem. For point-like
quarks they contain an $1/r^3$ term. Once again the finite size of the constituent 
quarks allows for a regularization, obtaining tensor and spin-orbit
potentials of the form,
\begin{eqnarray}
V_{OGE}^{T}(\vec{r}_{ij})& =&-{\frac{1}{16}}{\frac{\alpha_s}{{m_{i}m_{j}}}}
\vec{\lambda}_i^c \cdot \vec{\lambda}_j^c \left[ {1 \over r_{ij}^3} - 
{{e^{-r_{ij}/r_g(\mu)}} \over {r_{ij}}} \left( {1 \over r_{ij}^2} + 
{1 \over {3 r_g^2(\mu)}} + {1 \over {r_{ij} \, r_g(\mu)}} \right) \right] 
S_{ij}  \, , \nonumber \\
V_{OGE}^{SO}(\vec{r}_{ij}) & =& 
-{\frac{1}{16}}{\frac{\alpha_s}{{m_{i}^2m_{j}^2}}}
\vec{\lambda}_i^c \cdot \vec{\lambda}_j^c \left[{1 \over r_{ij}^3} - 
{{e^{-r_{ij}/r_g(\mu)}} \over {r_{ij}^3}} \left( 1 + 
{r_{ij} \over { r_g(\mu)}}  \right) \right] \, \times \, \\
&&\left[ \left( (m_{i}+m_{j})^{2}+2m_{i}m_{j}\right) (\vec{S}_{+}\cdot \vec{L%
})+\left( m_{j}^{2}-m_{i}^{2}\right) 
(\vec{S}_{-}\cdot \vec{L})\right] \, , \nonumber
\end{eqnarray}
where $\vec S_{\pm}=\vec S_i\pm\vec S_j$, and $r_g(\mu)=\hat r_g /\mu$
presents a similar behavior to the scaling of the central term.

The strong coupling constant, taken to be constant
for each flavor sector, has to be scale-dependent 
when describing different flavor sectors \cite{Tit95}. 
Such an effective scale dependence has been related to the typical
momentum scale of each flavor sector assimilated
to the reduced mass of the system \cite{Hal93}.
This has been found to be relevant for the study of the
meson spectra and parametrized in Ref. \cite{Vijb5} as
\begin{equation}
\alpha_s(\mu)={\alpha_0\over{ln\left[{({\mu^2+\mu^2_0})/
\gamma_0^2}\right]}},
\label{asf}
\end{equation}
where $\mu$ is the reduced mass of the interacting $qq$ 
pair and $\alpha_0$, $\mu_0$, and $\gamma_0$ are fitted parameters.

Finally, any model imitating QCD should incorporate
confinement. Lattice calculations
in the quenched approximation derived, for heavy quarks, a
confining interaction linearly dependent on the interquark
distance.  
The consideration of sea quarks apart from valence 
quarks (unquenched approximation) suggests a screening effect 
on the potential when increasing the interquark 
distance \cite{Bal01}. Creation of light-quark
pairs out of vacuum in between the quarks becomes energetically preferable
resulting in a complete screening of quark color charges at large distances.
String breaking has been definitively confirmed
through lattice calculations \cite{SESAM} 
in coincidence with the quite rapid
crossover from a linear rising to a flat potential well
established in $SU(2)$ Yang-Mills theories \cite{Est99}.
The central part of a screened potential simulating these results 
can be written as,
\begin{equation}
V^C_{CON}(\vec{r}_{ij})=-a_{c}\,(1-e^{-\mu_c\,r_{ij}})
(\vec{\lambda^c}_{i}\cdot \vec{ \lambda^c}_{j}) \, \, .
\end{equation}
At short distances it presents 
a linear behavior with an effective confinement strength 
$a=a_c \, \mu_c \, \vec{\lambda^c}_i \cdot \vec{\lambda^c}_j$, while
it becomes constant at large distances. Screened 
confining potentials have been analyzed in the literature
providing an explanation to the
missing state problem in the baryon spectra \cite{Vij03},
improving the description of the heavy-meson spectra \cite{Ped03},
and justifying the
deviation of the meson Regge trajectories from the linear behavior
for higher angular momentum states \cite{Bri00}.

Confinement also presents an spin-orbit contribution
taken to be an arbitrary combination of scalar and vector terms of the form
\begin{eqnarray}
V_{CON}^{SO}(\vec{r}_{ij}) &=&-(\vec{\lambda}_{i}^{c}\cdot
\vec{\lambda}_{j}^{c})\,{\frac{{a_c\,\mu_c 
\,e^{-\mu_c \,r_{ij}}}}{{4m_{i}^{2}m_{j}^{2}r_{ij}}}}
\left[ \left( (m_{i}^{2}+m_{j}^{2})(1-2\,a_{s}) \right. \right. \nonumber \\
&&\left. \left. +4m_{i}m_{j}(1-a_{s})\right) (\vec{S}_{+}\cdot \vec{L})+ 
(m_j^2-m_i^2) (1-2\,a_s)(\vec{S}_{-}\cdot \vec{L}) \right]
\end{eqnarray}
where $a_s$ would control the ratio between them.

Once perturbative (one-gluon exchange) and nonperturbative (confinement and
chiral symmetry breaking) aspects of QCD have been considered, one ends up with
a quark-quark interaction of the form 
\begin{equation} 
V_{q_iq_j}=\left\{ \begin{array}{ll} 
q_iq_j=nn/sn\Rightarrow V_{CON}+V_{OGE}+V_{\chi} &  \\ 
q_iq_j=cn/cs\Rightarrow V_{CON}+V_{OGE} & \end{array} \right.\,.
\label{pot}
\end{equation}
Note that for the particular case of heavy quarks, chiral symmetry is
explicitly broken and therefore Goldstone boson exchanges do not contribute.
The model parameters and a more detailed
discussion of the model can be found in Refs.~\cite{Vijb5,Vij05,Veu04}.

\section{RESULTS AND DISCUSSION}

A thoroughly study of the full meson spectra has been presented
in Ref. \cite{Vijb5}. The results for the open-charm mesons are 
resumed in Table \ref{t1}. It can be
seen how the open-charm states are easily identified with
standard $c \overline{q}$ mesons except for the cases of
the $D_{sJ}^*(2317)$, the $D_{sJ}(2460)$, and the $D^*_0(2308)$.
We will also comment on the recent measurement of the $D_1^0(2430)$. 
This behavior is shared by almost all quark potential model 
calculations \cite{God91}. 
Although the situation from lattice QCD is far from being definitively established,
similar problems are observed. Lattice NRQCD in the quenched 
approximation predicts for the $D_{sJ}^*(2317)$ a mass of 2.44 GeV \cite{Hei00}, 
while using relativistic charm quarks the mass obtained is 2.47 GeV \cite{Boy97}.
Unquenched lattice QCD calculations of $c \overline s$ 
states do not find a window for the $D^*_{sJ}(2317)$ \cite{Bal03},
supporting the difficulty of a $P-$wave $c\overline s$ interpretation. 
The quenched lattice QCD calculation of
the spectrum of orbitally excited $D_s$ mesons of Ref. \cite{Dou03} 
concludes that, although 
the results obtained are consistent with a $c\bar s$ configuration, the statistical and 
systematical uncertainties are too large to exclude the exotic states based on
potential quark models \cite{God91}.
The same situation may be drawn from heavy quark symmetry arguments. One finds
that the scalar $c\overline s$ state belongs to the $j=1/2$ doublet, 
but since the $j=3/2$ doublet is identified with the
narrow $D_{s2}(2573)$ and $D_{s1}(2536)$ (with total widths 
of 15$^{+5}_{-4}$ MeV and $<2.3$ MeV, respectively) the scalar state is
expected to have a much larger width than the one measured for the
$D_{sJ}^*(2317)$ \cite{WI90}. 

Thus, one could be tempted to interpret these states as four-quark resonances
within the quark model. The results obtained for the $cn\bar s\bar n$ 
and $cn\bar n\bar n$ configurations with the 
same interacting potential of Sec. II are shown in Table \ref{t2}. 
The $I=1$ and $I=0$ states are far above the corresponding 
strong decay threshold and therefore should be broad, what rules
out a pure four-quark interpretation of the new open-charm mesons. 

As outlined above, for $P-$wave mesons the hadronic dressing is in a $S-$wave,
thus physical states may correspond to a mixing of 
two- and four-body configurations, Eq. (\ref{mes-w}). 
In the isoscalar sector, the 
$cn\bar s \bar n$ and $c\bar s$ states get mixed, as it happens
with $cn \bar n\bar n$ and $c\bar n$ for the $I=1/2$ case. 
Let us notice that the interacting potential contains terms
mixing $P-$wave $S=0$ and $S=1$ states \cite{Vijb5}, arising from the 
different spin-orbit contributions for systems made
of quarks of different mass. The mixing angle comes determined
by the structure of the interaction.
The parameter $\gamma$ has been fixed to reproduce the 
mass of the $D_{sJ}^*(2317)$ meson, being $\gamma=240$ MeV. Using this value
one has $|\gamma/(E^{\,n}_{J^{PC}}-E^{\,n+1}_{J^{PC}})|^2\approx0.25\ll1$, a ratio
consistent with the assumption that the hamiltonian $H_1$, Eq. (\ref{eq1}),
can be treated perturbatively. This perturbative condition together with
$|\gamma/E_{q\bar q}|\approx|\gamma/E_{qq\bar q\bar q}|\approx0.1\ll1$ also ensures 
that the mixing with higher states can be neglected. 

The results obtained are shown in Table \ref{t3}. 
Let us first analyze the nonstrange sector. The 
$^{3}P_{0}$ $c\bar n$ pair and the $cn\bar n\bar n$ 
have a mass of 
2465 MeV and 2505 MeV, respectively. Once the mixing is considered
one obtains a state at 2241 MeV with 46\% of four-quark component 
and 53\% of $c\bar n$ pair. The lowest state, representing
the $D^*_0(2308)$, is above the isospin preserving threshold $D\pi$,
being broad as observed experimentally. 
The mixed configuration compares much better with 
the experimental data than the pure $c\bar n$ state. 
The orthogonal state appears higher in energy, at 2713 MeV, with
and important four-quark component. 
A similar process would modify the state representing
the $D_1^0(2430)$, but in this case one would need the mass of the $I=1/2$
$J^P=1^+$ four-quark state. The huge basis generated for such quantum numbers
makes unfeasible its calculation. A correct description of this state
would require a mass of 2.9 GeV for the four quark state mentioned above.

Concerning the strange sector, the $D_{sJ}^*(2317)$ and the $D_{sJ}(2460)$ 
are dominantly $c\bar s$ $J=0^+$ and $J=1^+$ states, respectively,
with almost  30\% of four-quark component. Without being dominant,
it is fundamental to shift the mass of the unmixed states to 
the experimental values below the $DK$ and $D^*K$ thresholds.
Being both states below their isospin-preserving 
two-meson threshold, the only allowed strong decays to
$D_s^* \pi$ would violate isospin and are expected to
have small widths. This width has been estimated assuming either a $q\bar q$ structure 
\cite{God03,Bar03}, a four-quark state \cite{Nie05} or vector meson dominance \cite{Col03} 
obtaining in all cases a width of the order of 10 keV.
The second isoscalar $J^P=1^+$ state, with an energy of 2555 MeV and 
98\% of $c\bar{s}$ component, corresponds to the $D_{s1}(2536)$. 
Regarding the $D_{sJ}^*(2317)$, it has been argued that a 
possible $DK$ molecule would be preferred with
respect to an $I=0$ $cn\bar s\bar n$ tetraquark, 
what would anticipate an $I=1$ $cn\bar s\bar n$ partner 
nearby in mass \cite{Barb3}. Our results confirm the last argument, 
the vicinity of the isoscalar and isovector tetraquarks (see Table \ref{t2}),
however, the restricted coupling to the $c\bar s$ system allowed only for the
$I=0$ four-quark states opens the possibility of a mixed nature for the
$D_{sJ}^*(2317)$, the $I=1$ tetraquark partner remaining much higher in energy.
The $I=1$ $J=0^{+}$ and $J=1^{+}$ four-quark states appear above
2700 MeV and cannot be shifted to lower energies.

Our results do not show any state 
around 2600 MeV, the mass region compatible with the 
$D_{sJ}^+(2632)$ measured by SELEX. Its two possible theoretical 
partners, the $2^{3}S_{1}$ state (suggested in Ref. \cite{Bev03})
or the $1^{3}D_{1}$ state, lie above 2700 MeV (2764 and 2873 MeV, 
respectively). This state has been proposed as the
$cs\bar s\bar s$ partner of the $D_{sJ}^*(2317)$ \cite{Cheb3}. 
Naively, the $D_{sJ}^*(2317)$ sets a scale of 2320 MeV for 
the $cn\bar s\bar n$ sector, 
and an augment of 150 MeV for each strange quark
could accommodate a $cs\bar s\bar s$ system near the
mass of the $D_{sJ}^+(2632)$. However, the mass of the 
$D_{sJ}^*(2317)$ is obtained through the coupling 
to the $c\bar{s}$ quark pair, being the mass of the tetraquark
configuration above 2700 MeV, disregarding a possible $cs\bar s\bar s$
interpretation of the $D_{sJ}^+(2632)$. It has also been discarded as the first
radial excitation of the $D^*_s(2112)$ \cite{Cha05}. A careful
analysis of several theoretical interpretations has been done in
Ref. \cite{Bar04}, the surprising properties of this state not
fitting in any of the scenarios considered. Several experiments
have failed trying to confirm the existence of this state \cite{Bab04}.
The confirmation or
refutation of the $D_{sJ}^+(2632)$ is clearly an important
priority for meson spectroscopy.

The structure of the $D^*_{sJ}(2317)$ and the $D_{sJ}(2460)$ mesons could 
be scrutiny, apart from their masses, also through the study
of their electromagnetic decay widths. 
Using the standard formalism described, for example, in Ref. \cite{God03}, the
$E1$ radiative transitions for $c\bar s \to c\bar s+\gamma$ are given by:
\begin{equation}
\Gamma[c\bar s \to c\bar s+ \gamma]={4\over27}\alpha\left<e_Q\right>^2\omega^3(2J_f+1)
\left|\left<^{2S+1}S_{J'}\left|r\right|^{2S+1}P_{J}\right>\right|^2S_{if}
\end{equation}
where $S_{if}$ is a statistical factor with $S_{if}=1$ for the transitions
between spin-triplet states and $S_{if}=3$ for the transition between
spin-singlet states, $\left<e_Q\right>$ is an effective quark charge given by
\begin{equation}
\left<e_Q\right>={{m_se_c-m_ce_{\bar s}}\over{m_c+m_s}}
\end{equation}
where $e_i$ is the charge of the quark(antiquark) $i$ in units of $|e|$,
$\alpha$ is the fine structure constant and $\omega$ is the photon energy.
Once the mixing between two- and four-quark components has been included there
would also be a contribution to the decay width coming from 
$cn\bar s\bar n\to c \bar s+\gamma$, that goes necessarily through the
annihilation of a color singlet $n\bar n$ pair with photon quantum numbers
within the four-quark wave function. The contribution to the electromagnetic
decay width arising from the four-quark component gets suppressed. This can be
illustrated by the analysis of the $M1$ electromagnetic decay width for 
$cn\bar s\bar n(1^+)\to cn\bar s\bar n(0^+)+\gamma$, that is given by
\begin{equation}
\Gamma[cn\bar s\bar n(1^+)\to cn\bar s\bar n(0^+)+\gamma]
={4\over3}\alpha\omega^3(2J_f+1)
\left|\left<^1S_0\left|\mu_z\right|^{3}S_{1}\right>\right|^2
\end{equation}
where
\begin{equation}
\mu_z=\sum_{i=1}^4{{e_i}\over{2m_i}}\sigma_{iz}\,.
\end{equation}
From the above equation one obtains $\Gamma[cn\bar s\bar n(1^+)\to cn\bar s\bar n(0^+)+\gamma]\approx 4$ eV,
that will contribute to the $\Gamma[D_{sJ}(2460)\to {D^*_{sJ}(2317)+\gamma}]$ 
decay width, once the configuration probabilities given in Table \ref{t3} are considered, 
in the order of 0.25 eV. The value obtained in Ref. \cite{God03} for this decay, assuming 
a pure $c\bar s$ configuration, is one order of magnitude larger confirming the smallness
of the four to four-quark contribution. In Ref. \cite{Clo05} this
decay width was estimated assuming a molecular structure, obtaining a
larger value of the order of 17 keV. This makes evident the important 
difference between the assumed molecular structure of Ref. \cite{Clo05} and the full multiquark 
configuration considered in this work. In the case of the
four-quark configuration the orthogonality between the several components of
the wave functions diminishes the decay widths, an effect that does not seem to be present 
within the molecular interpretation. 
The same reasoning applies to the $\Gamma[cn\bar s\bar n\to c\bar s+\gamma]$
decay width, being suppressed due to the reduced singlet-singlet component with
photon quantum numbers within the four-quark wave function.
Similar conclusions were obtained in Ref. \cite{Bro04}. As a consequence,
the presence of a four-quark
component diminishes the decay widths, making them different
from those predicted for a pure $c \bar s$ state \cite{God03,Barb3,Col03}. 

We compare in Table \ref{t4} our results for the radiative transitions of the 
$D^*_{sJ}(2317)$ and $D_{sJ}(2460)$ with different theoretical 
approaches and the experimental limits reported by CLEO and Belle. 
The main difference is noticed in the suppression predicted 
for the $D_{sJ}(2460)\to D^{*+}_s\gamma$ decay as 
compared to the $D_{sJ}(2460)\to D^{+}_s\gamma$. 
A ratio ${D_{sJ}(2460)\to D^{+}_s\gamma}/{D_{sJ}(2460)
\to D^{*+}_s\gamma}\approx1-2$ has been obtained 
assuming a $q\bar q$ structure for both states \cite{God03,Bar03}
(what seems incompatible with their properties).
Heavy-hadron chiral perturbation theory calculations find a similar 
ratio \cite{Meh04}. We find a larger value,
${D_{sJ}(2460)\to D^{+}_s\gamma}/{D_{sJ}(2460)
\to D^{*+}_s\gamma}\approx100$, due to the small
$1^3P_1$ $c\overline s$ probability of the 
$D_{sJ}(2460)$. A similar enhancement 
has been obtained in Ref. \cite{Col05} (see 
penultimate column of Table \ref{t4}) in the
framework of light-cone QCD sum rules in contrast to
a previous calculation of the same authors
using vector meson dominance \cite{Col03}. 
As a consequence, the radiative transitions
are an important diagnostic tool to understand the nature
of these states. In view of the different predictions of the electromagnetic decay widths,
a precise measurement of this decay would allow to distinguish not 
only between $q\bar q$ and non$-q\bar q$ states, but also between pure molecular and four-quark 
interpretations.

Let us finally mention that the difficulties encountered for the interpretation
of the new open-charm states as two-quark systems do not appear for the case of
the recent charmed and $B_c$ states measured at different facilities. They do
nicely fit into the predictions of the model used, see Table \ref{t5},
giving confidence to the results obtained in the present work.

The interpretation we have just presented of the positive parity open-charmed mesons as a mixture
of two- and four-quark states has also been used to account for the experimentally observed light-scalar mesons within the
same constituent quark model \cite{Vij05},
what gives us confidence on the mechanism proposed. Nonetheless, one should not
forget that in the literature there is a wide variety of interpretations for
the open-charm and also for the light-scalar mesons. As previously mentioned, they
used to deal with one of both problems and even then with a particular set of
states, being the most common ones the strange open-charm states and the
isoscalar light mesons.

Regarding the light-scalar mesons, there have been reported by the PDG two
isovectors, five isoscalar and three isodoublet states. The situation is far
from being definitively settled, neither from the experimental nor from the
theoretical point of view. Experimentally, let us only mention that the recent
analysis by BES of the $J/\Psi \to \phi \pi^+ \pi^-$ and $J/\Psi \to \phi K^+ K^-$ data \cite{Ab04b}
requires a state $f_0(1790)$ distinct from the $f_0(1710)$. Recent reanalysis of the Crystal Barrel data
\cite{AS02} suggest the existence of a new state called $f_0(1200-1600)$.
Theoretically, one would expect non$-q \bar q$ scalar objects in the mass range below 2 GeV.
Multiquarks have been justified to coexist with $q\bar q$ states in the energy region
around 1 GeV because they can couple to $0^{++}$ without
orbital excitation \cite{Jaf77}. Lattice QCD in the quenched approximation 
predicts the existence of a scalar glueball with a mass around 1.6 GeV \cite{Mi97}. 

From this complicated scenario many different interpretations of the
light-scalar mesons have arisen.
The overpopulation of $0^{++}$ states below 2 GeV gave rise long
ago to the speculation of the existence of four light-quark states. The most complete 
analysis was performed by Weinstein and Isgur \cite{Wei83,WI82}, concluding that, normally,
$qq\bar q\bar q$ bound states do not exist, being the only exception the scalar sector 
where weakly bound states with a meson-meson molecule structure were found.  
Particular sets of states have been studied in the literature. There are several 
models analyzing the mixing between
different configurations to yield the physical $f_0(1370)$,
$f_0(1500)$, and $f_0(1710)$. Among them, Ref. \cite{Vij05} assigning the 
larger glueball component to the $f_0(1710)$ is on the line with 
Refs. \cite{Mcn00} and differ from those of Refs. \cite{Ams02,Clo05b} 
concluding that the $f_0(1710)$ is dominantly $q \bar q$. One should notice that in these 
studies only Refs. \cite{Vij05,Clo05b} consider the recently reported $f_0(1790)$.
In a different fashion within the quark model, the $a_0(980)$ and $f_0(980)$ mesons were analyzed
in Ref. \cite{TORN}, being the effect of the two-pseudoscalar meson thresholds 
the responsible for the substantial shift
to a lower mass than what is naively expected from the $q\bar q$ component alone.
This gives rise to an important $K\overline K$ and $\pi\eta'$ components in the
$a_0(980)$ and $K\overline K$,  $\eta\eta$, $\eta'\eta'$ and $\eta\eta'$ in the
$f_0(980)$. 

The structure of the scalar mesons $a_0(980)$ and $f_0(980)$ have been also investigated in
the framework of a meson exchange based on the J\"ulich potential model \cite{PRD52}
for $\pi\pi$ and $\pi\eta$ scattering. Whereas the $f_0(980)$
appears to be a $K\overline K$ bound state the $a_0(980)$ was found to be a dynamically
generated threshold effect. Similar conclusions have been obtained in a
chiral unitary coupled channel approach, where the $f_0(600)$, the $a_0(980)$, and the 
$K^*_0(800)$ rise up as dynamically generated resonances, while the $f_0(980)$ is a 
combination of a strong $S-$wave meson-meson unitarity effect and a preexisting 
singlet resonance \cite{Ose97}.  In Ref. \cite{Bev00} van Beveren {\it et al.} describe 
the light scalar mesons as resonances and bound states
characterized by complex singularities of the scattering amplitude.

Finally, let us stress that Ref. \cite{Vij05} presents an interpretation
of the scalar mesons in a model constrained by the description of other hadron
sectors. The same mechanism has been applied here to disentangle the  structure
of the new $D_s$ and $D_{sJ}$ resonances. It drives to a final scenario that 
it is compatible with some other models in the literature and it differs from 
other results.
Being the set of data so huge, and sometimes so poor, one always 
may find a positive or negative interpretation of some of them. 
Therefore, the final answer could only be obtained from precise experimental 
data that would allow to discriminate between the predictions of different 
theoretical models \cite{Ams04}.

\section{SUMMARY}

As a summary, we have obtained a rather satisfactory description of 
the positive parity open-charm mesons in terms of two- and four-quark
configurations. The mixing between these two components is
responsible for the unexpected low mass and widths of the
$D_{sJ}^*(2317)$, $D_{sJ}(2460)$, and $D_0^*(2308)$. 
The same mechanism has been used to account for the spectroscopic properties of
the light-scalar mesons.
The obtained electromagnetic decay widths give hints that would help 
in distinguishing the nature of these states.
We predict a ratio ${D_{sJ}(2460)\to
D^{+}_s\gamma}/{D_{sJ}(2460)\to D^{*+}_s\gamma}$ much larger
than the one obtained in a pure $q\bar q$ scheme.
We did not find any theoretical partner for the recently
measured $D_{sJ}^+(2632)$ whose existence awaits 
confirmation \cite{Bab04}. We encourage experimentalists
on two different directions: the measurement of the electromagnetic
decay widths of the $D^*_{sJ}(2317)$ and the $D_{sJ}(2460)$,
and the confirmation or refutation of the 
$D_{sJ}^+(2632)$ that would help to clarify the exciting
situation of the open-charm mesons.

\section{acknowledgments}
This work has been partially funded by Ministerio de Ciencia y Tecnolog\'{\i}a
under Contract No. FPA2004-05616, by Junta de Castilla y Le\'{o}n
under Contract No. SA-104/04, and by Generalitat Valenciana under Contract No.
GV05/276.

\newpage

\begin{table}
\caption{$c\overline s$ and $c\overline n$ masses (QM), in MeV.
Experimental data (Exp.) are taken from
Ref. \protect\cite{Eid04}, except for the state denoted by a dagger
that has been taken from Ref. \protect\cite{Belb4}.}
\label{t1}
\begin{center}
\begin{tabular}{|cc|ccc||ccc|}
\hline
&$nL$ $J^P$&State		  &QM $(c\overline s)$  &Exp.				&State			&QM $(c\overline n)$ 	&Exp.\\
\hline
&$1S$ $0^-$&$D_{s}$ 	  & 1981  &1968.5$\pm$0.6	&$D$			&1883 	&1867.7$\pm$0.5\\
&$1S$ $1^-$&$D^*_{s}$	  & 2112  &2112.4$\pm$0.7	&$D^*$			&2010 	&2008.9$\pm$0.5\\
&$1P$ $0^+$&$D^*_{sJ}(2317)$& 2489  &2317.4$\pm$0.9	&$D^*_0(2308)$	&2465 	&2308$\pm$17$\pm$15$\pm$28$^\dagger$\\
&$1P$ $1^+$&$D_{sJ}(2460)$& 2578  &2459.3$\pm$1.3	&$D_1(2420)$	&2450 	&2422.2$\pm$1.8\\
&$1P$ $1^+$&$D_{s1}(2536)$& 2543  &2535.3$\pm$0.6	&$D_1^0(2430)$	&2546 	&$2427\pm26\pm25$\\
&$1P$ $2^+$&$D_{s2}(2573)$& 2582  &2572.4$\pm$1.5	&$D_2^*(2460)$	&2496 	&2459$\pm$4\\
\hline
\end{tabular}
\end{center}
\end{table}

\begin{table}
\caption{$cn\bar s\bar n$ and $cn\bar n\bar n$ masses, in MeV.}
\label{t2}
\begin{center}
\begin{tabular}{|cc|cc|c|}
\hline
\multicolumn{4}{|c|} {$cn\bar s\bar n$} & $cn\bar n\bar n$ \\
\hline
\multicolumn{2}{|c|} {$J^{P}=0^{+}$} & \multicolumn{2}{|c|}{$J^{P}=1^{+}$}&$J^{P}=0^{+}$ \\
$I=0$& $I=1$ &  $ I=0$& $I=1$  & $I=1/2$\\
\hline
2731&  2699& 2841&2793 &2505\\
\hline
\end{tabular}
\end{center}
\end{table}

\begin{table}
\caption{Probabilities (P), in \%, of the wave function components 
and masses (QM), in MeV, of the open-charm mesons 
once the mixing between $q\bar q$ and $qq\bar q\bar q$ configurations 
is considered. Experimental data (Exp.) are taken from Ref. \cite{Eid04} 
except for the state denoted by a dagger that has been taken from
Ref. \protect\cite{Belb4}.} 
\label{t3}
\begin{center}
\begin{tabular}{|c|cc||c|cc||c|cc|}
\hline
\multicolumn{6}{|c||}{$I=0$} & \multicolumn{3}{|c|}{$I=1/2$} \\
\hline
\multicolumn{3}{|c||}{$J^P=0^+$}    & \multicolumn{3}{|c||}{$J^P=1^+$} & \multicolumn{3}{|c|}{$J^P=0^+$} \\
\hline
QM                  &2339   &2847  &QM					&2421  &2555  	&QM                   &2241 &2713    \\
Exp.                &2317.4$\pm$0.9&$-$  &Exp.		&2459.3$\pm$1.3&2535.3$\pm$0.6&Exp.   &2308$\pm$17$\pm$15$\pm$28$^\dagger$&$-$\\
\hline
P($cn\bar s\bar n$) &28   &55  &P($cn\bar s\bar n$)	&25  &$\sim 1$ 	&P($cn\bar n\bar n$)  &46        &49  \\
P($c\bar s_{1^3P}$) &71   &25  &P($c\bar s_{1^1P}$)	&74  &$\sim 1$ 	&P($c\bar n_{1P}$)    &53        &46 \\
P($c\bar s_{2^3P}$) &$\sim 1$  &20  &P($c\bar s_{1^3P}$)&$\sim 1$ &98	&P($c\bar n_{2P}$)    &$\sim 1$  &5 \\
\hline
\end{tabular}
\end{center}
\end{table}

\begin{table}
\caption{Comparison of 90\% C.L. limits on radiative transitions
obtained by CLEO \protect\cite{Cle03} and Belle \protect\cite{Bel04}
with our results (QM) and those of two different 
quark models, Refs. \cite{Bar03,God03}, based only on $q\overline q$ 
components. The BR's are with respect to the decay
$D^*_{sJ}(2317) \to D_s^+ \pi^0 $ for the
$D^*_{sJ}(2317)$ and with respect to the decay
$D_{sJ}(2460) \to D_s^{*+} \pi^0 $ for the
$D_{sJ}(2460)$. We have assumed 
$\Gamma( D^*_{sJ}(2317) \to D_s^+ \pi^0) \approx
\Gamma( D_{sJ}(2460) \to D_s^{*+} \pi^0)
\approx 10$ keV as explained in the text.
We have also quoted results obtained by 
light-cone QCD sum rules \protect\cite{Col05} and
vector meson dominance \protect\cite{Col03}.}
\label{t4}
\begin{center}
\begin{tabular}{|c|ccc|cc|cc|}
\hline
 & \multicolumn{3}{|c|}{Quark models} & \multicolumn{2}{|c|}{Experiments} & \multicolumn{2}{|c|}{Other approaches} \\
Transition & QM &Ref. \cite{Bar03} &Ref. \cite{God03} &CLEO \cite{Cle03}& Belle \cite{Bel04} &
Ref. \cite{Col05} & Ref. \cite{Col03} \\
\hline
${D^*_{sJ}(2317)\to D^{*+}_s\gamma}$ & 0.16  & 0.17  & 0.19 & $<0.059$ & $<0.18$ & $0.4-0.6$     & 0.085   \\
${D^*_{sJ}(2317)\to D^{+}_s\gamma}$  & 0.0  & 0.0  & 0.0  & $<0.052$ & $<0.05$ & 0.0       & 0.0    \\
${D_{sJ}(2460)\to D^{*+}_s\gamma}$   & 0.006 & 0.47  & 0.55 & $<0.16$  & $<0.31$ & $0.06-0.11$ & 0.15    \\
${D_{sJ}(2460)\to D^{+}_s\gamma}$    & 0.67  & 0.51  & 0.62 & $<0.49$  &0.55$\pm$0.13$\pm0.08$ &  $1.9-2.9$ & 0.33\\
\hline
\end{tabular}
\end{center}
\end{table}

\begin{table}
\caption{Masses (QM), in MeV, of the recently 
measured charmonium and $B_c$ 
states obtained within the model of Ref. \protect\cite{Vijb5} used in this
work.}
\label{t5}
\begin{center}
\begin{tabular}{|ccc|cc|}
\hline
Name & Mass & Ref. & $n^{2S+1}L_J$ & QM \\
\hline
$X(3940)$        & $3943\pm 6\pm 6$   & \cite{Bel05} & $2^1P_1$   & 3923 \\
$-      $        & $-             $   &              & $2^3P_0$   & 3878 \\
$Y(3940)$        & $3943\pm 11\pm 13$ & \cite{Belb5} & $2^3P_1$   & 3915 \\
$X'_{c2}(3940)$  & $3931\pm 4\pm 2$   & \cite{Belc5} & $2^3P_2$   & 3936 \\
$Y(4260)$        & $4260\pm 8\pm 2$   & \cite{Beld5} & $4^3S_1$   & 4307 \\
$B_c(6287)$      & $6287\pm4.8\pm1.1$ & \cite{Cdf05} & $1^1S_0$   & 6277 \\
\hline
\end{tabular}
\end{center}
\end{table}

\end{document}